\documentclass[reprint,aps,prl,superscriptaddress,showpacs]{revtex4-1}

\usepackage{amssymb,amsmath,amsfonts}
\usepackage{graphicx}
\usepackage{xcolor}
\usepackage{hyperref}
\hypersetup{
    colorlinks,
    linkcolor={blue!60!black},
    citecolor={blue!60!black},
    urlcolor={blue!60!black}
}

\begin{document}

\title{Time-Resolved Observation of Thermalization in an Isolated Quantum System} 

\author{Govinda Clos}
\email{govinda.clos@physik.uni-freiburg.de}
\affiliation{Physikalisches Institut, Albert-Ludwigs-Universit{\"a}t, Hermann-Herder-Stra{\ss}e 3, 79104 Freiburg, Germany.}
\author{Diego Porras}
\affiliation{Department of Physics and Astronomy, University of Sussex, Brighton BN1 9QH, United Kingdom.}
\author{Ulrich Warring}
\affiliation{Physikalisches Institut, Albert-Ludwigs-Universit{\"a}t, Hermann-Herder-Stra{\ss}e 3, 79104 Freiburg, Germany.}
\author{Tobias Schaetz}
\affiliation{Physikalisches Institut, Albert-Ludwigs-Universit{\"a}t, Hermann-Herder-Stra{\ss}e 3, 79104 Freiburg, Germany.}

\date{\today}

\begin{abstract}
We use trapped  atomic ions forming a hybrid Coulomb crystal, and exploit its phonons to study an isolated quantum system composed of a single spin coupled to an engineered bosonic environment.
We increase the complexity of the system by adding ions and controlling coherent couplings and, thereby, we observe the emergence of thermalization:
Time averages of spin observables approach microcanonical averages while related fluctuations decay.
Our platform features precise control of system size, coupling strength, and isolation from the external world to explore the dynamics of equilibration and  thermalization.
\end{abstract}
\maketitle

How does statistical mechanics emerge from the microscopic laws of nature? Consider, for example, a finite, isolated quantum system: It features a discrete spectrum and a quantized phase space, its dynamics are governed by the linear Schr\"odinger equation and, thus, remain reversible at all times. Can such a system equilibrate or even thermalize? Progress in the theory of nonequilibrium dynamics and statistical mechanics  sheds light on these fundamental questions. It has been shown that individual quantum states can exhibit properties of thermodynamics depending on entanglement within the system~\cite{popescu_entanglement_2006,goldstein_canonical_2006,reimann_foundation_2008,linden_quantum_2009,cramer_quantum_2010,short_quantum_2012,riera_thermalization_2012}. While the entire system may very well be described by a pure state, any small subsystem and related local observables may be found in a mixed state due to disregarded entanglement with the rest of the isolated system, i.e., the large environment. Further, it is predicted that even any individual many-body eigenstate of a nonintegrable Hamiltonian yields expectation values for few-body observables that are indistinguishable from microcanonical averages~\cite{deutsch_quantum_1991,srednicki_chaos_1994,srednicki_approach_1999,rigol_thermalization_2008,polkovnikov_colloquium:_2011,dalessio_quantum_2016}. This conjecture has been extensively studied by numerical simulations of specific quantum many-body systems of moderate size, exploiting available computational power~\cite{rigol_breakdown_2009,santos_onset_2010,beugeling_finite-size_2014,kim_testing_2014}.  Recently, there have been first experiments in the context of thermalization in closed quantum systems with ultracold atoms \cite{gring_relaxation_2012,trotzky_probing_2012,langen_experimental_2015}. 
However, fundamental questions on the underlying microscopic dynamics of thermalization and its time scales remain unsettled~\cite{cazalilla_focus_2010,polkovnikov_colloquium:_2011,eisert_quantum_2015}.

Trapped-ion systems are well suited to study quantum dynamics at a fundamental level, featuring unique control in preparation, manipulation, and detection of electronic and motional degrees of freedom\,\cite{wineland_experimental_1998,leibfried_quantum_2003,blatt_entangled_2008,myerson_high-fidelity_2008,kienzler_quantum_2015,tan_multi-element_2015,ballance_hybrid_2015}. Their Coulomb interaction of long range permits tuning from weak to strong coupling \cite{porras_effective_2004}. 
Additionally, systems can be scaled bottom up to the mesoscopic
size of interest to investigate many-body physics \cite{schneider_experimental_2012,monroe_scaling_2013,ramm_energy_2014,mielenz_arrays_2016}.

In this Letter, we study linear chains of up to five trapped ions using two different isotopes of magnesium to realize a single spin with tunable coupling to a resizable bosonic environment.
Time averages of spin observables become indistinguishable from microcanonical ensemble averages and amplitudes of time fluctuations decay as the effective system size is increased.
We observe the emergence of statistical mechanics in a near-perfectly-isolated quantum system, despite its seemingly small size. 

The dynamics of our system are governed by the Hamiltonian~\cite{porras_mesoscopic_2008,supplement}
\nocite{braak_integrability_2011}
\nocite{caux_remarks_2011}
\nocite{weiss_quantum_1999}
\nocite{luitz_many-body_2015}
\nocite{mondaini_eigenstate_2016}
\nocite{hume_two-species_2010}
\nocite{friedenauer_high_2006}
\nocite{ansbacher_precision_1989}
\nocite{ozeri_errors_2007}
\nocite{casati_band-random-matrix_1993}
\begin{equation}
H =
\frac{\hbar\omega_{{z}}}{2} \sigma_z +
\frac{\hbar\Omega}{2} \sigma_x +
\sum_{j=1}^N \hbar\omega_j a^\dagger_j a_j+
\frac{\hbar\Omega}{2}\left(\sigma^+C    +\sigma^-C^\dagger \right).
\label{fullham}
\end{equation}
The spin is described by Pauli operators $\sigma_l\; (l=x,y,z)$ and $\hbar$ denotes the reduced Planck constant. The first term can be interpreted as interaction of the spin with an effective magnetic field $\omega_{{z}}$, lifting degeneracy of the eigenstates of
$\sigma_z$, labeled $|{\downarrow}\rangle$ and $|{\uparrow}\rangle$, while the second drives
oscillations between these states with spin coupling rate $\Omega$.
The sum represents the environment composed of $N$ harmonic oscillators with incommensurate frequencies $\omega_j$, and the operators $a_j$ $(a_j^\dagger)$ annihilate (create) excitations, i.e., phonons, of mode $j$.
The last term describes spin-phonon coupling via spin flips, $\sigma^{\pm}\equiv(\sigma_x\pm\text{i}\sigma_y)/2$, accompanied by motional (de)excitation which is incorporated in 
\begin{equation}
  \label{eq:1}
C = {\text{exp}}\left[{i \sum_{j=1}^N \eta_j \left(a_j^\dagger + a_j\right)}\right] - 1,  
\end{equation}
at a strength tunable by $\Omega$ and the spin-phonon coupling parameters $\eta_j \propto 1/\sqrt{\omega_j}$.
Expanding $C$ in series permits restricting to linear terms for values $\eta_j \ll 1$ (weak coupling). For $\eta_j \approx 1$, as in our experiment, higher order terms become significant (strong coupling), allowing the system to explore the full many-body set of highly entangled spin-phonon states. 
This regime is well described by full exact diagonalization (ED) only, since the discrete nature of the bosonic environment of finite size hinders standard approximations applicable to the spin-boson model considering a continuous spectral density~\cite{leggett_dynamics_1987,supplement}.

To study  nonequilibrium dynamics of  expectation values $\langle \sigma_l(t)\rangle\;(l=x,y,z)$,
consider an initial product state $\rho(t{=}0) \equiv \rho(0) = \rho_{{S}}(0) \otimes \rho_{{E}}(0)$, where the spin is in a pure
excited state, and the bosonic modes are cooled close to their motional ground states (average occupation  $\bar{n}_{j=1\ldots N} \lesssim 1$). 
With this, we ensure that energies of  spin and  phonons remain comparable
 to enable the observation of the coherent quantum nature of the dynamics
which creates entanglement of spin and phonon degrees of freedom. 
Because of the coupling, the spin subsystem is in a mixed state for $t>0$, even though the entire system is evolving unitarily.
When thermalization occurs, any small subsystem of a large isolated system equilibrates towards a thermal state and remains close to it for most times~\cite{srednicki_approach_1999,linden_quantum_2009}.

The so-called eigenstate thermalization hypothesis provides a potential explanation for the emergence of thermalization in an isolated quantum system. It can be phrased as a statement about matrix elements of few-body observables in the eigenstate basis of a many-body Hamiltonian~\cite{deutsch_quantum_1991,srednicki_chaos_1994,srednicki_approach_1999,rigol_thermalization_2008,dalessio_quantum_2016,polkovnikov_colloquium:_2011}.
Within this conjecture, infinite-time averages of expectation values of these observables  agree with microcanonical averages.
A  mathematical definition of this hypothesis and further information are given in the Supplemental Material.
Based on Refs.~\cite{deutsch_quantum_1991,srednicki_chaos_1994},
to interpret experimental results,
 we assume that a coupling distributes any of the energy eigenstates of an uncoupled system $\{ | \phi_\alpha \rangle \}$ over a large subset of the energy eigenstates of the coupled system $\left\{|\psi_\beta \rangle\right\}$, i.e., $| \phi_\alpha \rangle = \sum_{\beta} c_\beta(\alpha) | \psi_\beta \rangle$ \cite{supplement}. Further, we consider that these participating states lie within a narrow energy shell around the energy of $| \phi_\alpha \rangle$ \cite{rigol_thermalization_2008,dalessio_quantum_2016}.
As introduced in Refs.~\cite{srednicki_approach_1999,popescu_entanglement_2006,linden_quantum_2009,short_quantum_2012}, an effective dimension of the subset,  $d_\text{eff}\equiv 1/\text{tr}(\rho^2)$, provides an estimation for  the ergodicity of a system. 
It has been shown that mean amplitudes of time fluctuations of expectation values are bounded by $1 /  \sqrt{{d}_\text{eff}}$~\cite{linden_quantum_2009,short_quantum_2012}.

\begin{figure}[t]
  \centering
  \includegraphics[]{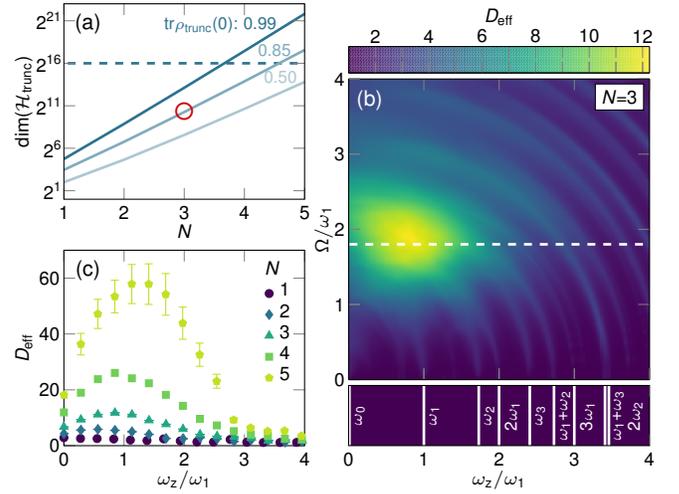}
  \caption{\label{fig1}Complexity of the Hamiltonian studied numerically. Parameters are $\omega_1/(2\pi)= 0.7\;$MHz, $\bar{n}_{j=1\ldots N}=1$.
(a) Dimension of truncated Hilbert space ${\text{dim}}({\mathcal H}_\text{trunc}(N))$ for corresponding fractions of initial-state population ${\text{tr}} \rho_\text{trunc}(0)$ lying within ${\mathcal H}_\text{trunc}$ (solid lines). 
For $N=3$, for example, 85\% lie within  ${\text{dim}}({\mathcal H}_\text{trunc})\approx 2^{10}$ (circle).
We derive $D_{\text{eff}}(N,\Omega,\omega_{{z}})$ up to ${\text{dim}}(\mathcal{H}_\text{trunc})= 2^{16}$ (dashed line).
(b) Choosing $\Omega$ and varying $\omega_{{z}}$ (dashed line), we can tune the spin-phonon coupling into resonance with different modes (sketched at the bottom) and boost the system size. 
Note, that the actual number of participating states is much larger than the normalized quantity $D_\text{eff}$; see Eq.~(\ref{deff}).
(c) For fixed $\Omega(N)$ [cf.~dashed line in (b)],  $D_{\text{eff}}(\omega_{{z}})$ increases significantly with $N$.
This enables the systematic investigation of equilibration and thermalization depending on the system size.
Error bars show systematic numerical uncertainties  \cite{supplement}.}
\end{figure}

Correspondingly, for our system, we exploit these  predictions for infinite-time averages, both of spin expectation values,
\begin{equation}
  \label{eq:2}
  {\mu_{\infty}}(\langle\sigma_l\rangle) \equiv \lim_{\tau \to \infty} \frac{1}{\tau} 
\int_0^\tau dt  \langle\sigma_l(t)\rangle
\end{equation}
  and of their time fluctuations,
  \begin{equation}
    \label{eq:fluct}
     {\delta_{\infty}}(\langle\sigma_l\rangle) \equiv \sqrt{{\mu_{\infty}}(\langle\sigma_l\rangle^2)-{\mu_{\infty}}(\langle\sigma_l\rangle)^2}.
  \end{equation}
To this end, we need to quantify the complexity of the dynamics induced by the coupling. Hence, we extend existing definitions of $d_\text{eff}$ to a weighted effective dimension \cite{supplement}
\begin{equation}
D_{\text{eff}} \equiv \sum_\alpha w_\alpha \bigg(\sum_\beta |c_\beta(\alpha)|^4\bigg)^{-1},
\label{deff}
\end{equation}
for $\rho(0)=\sum_\alpha w_\alpha | \phi_\alpha\rangle \langle\phi_\alpha|$. Here, in contrast to $d_\text{eff}$, the statistical average over $w_\alpha$ is performed after calculating the inverse participation ratio for each pure state in the mixture \cite{supplement}. Thereby, $D_\text{eff}$ also incorporates the  number of participating states, but is normalized to $D_\text{eff}=1$ for the uncoupled system.

\begin{figure*}[t]
  \centering
  \includegraphics[]{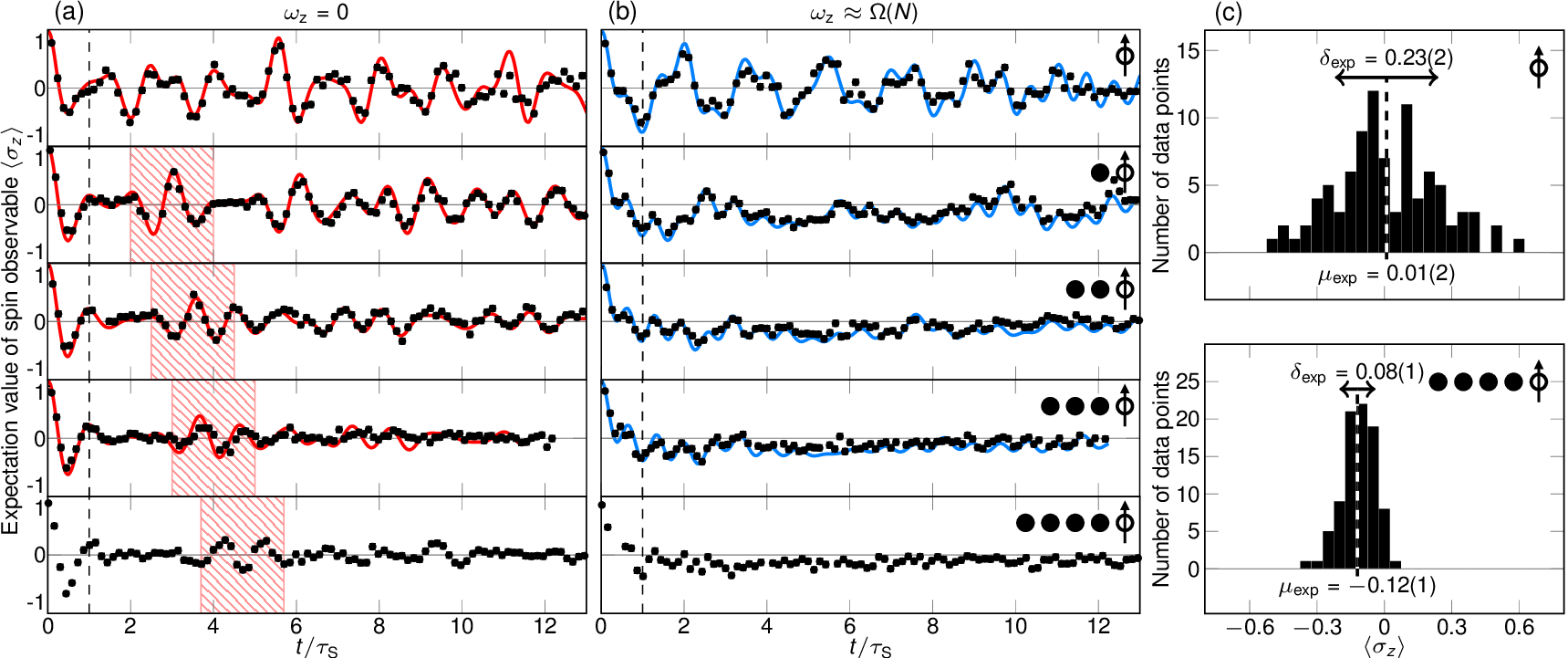}
  \caption{\label{fig2}Measured unitary time evolution $\langle \sigma_z(t) \rangle$. 
Experimental results (black dots, error bars: 1 s.d.)  for $N=1\ldots5$  compared to full ED (solid lines). We exclude numerical results for $N=5$ due to their large systematic uncertainties.
Oscillations (time fluctuations) of high amplitude during the transient duration $t/\tau_{{S}} \lesssim 1$, are driven by the evolution of $\rho(0)$ towards the ground state of $H$.
(a) For $\omega_{z}=0$ and increasing $N$, excitation is coherently exchanged with a growing number of modes resulting in revivals at $\tau_{\text{rev}}$ (shaded areas).
(b) For  $\omega_{z}\approx \Omega(N)$, expectation values fluctuate around a negative offset. Revivals and this nontrivial bias emphasize the coherence of the dynamics.
(c) Histograms of experimental measurements sample the probability distribution which underlies $\langle \sigma_z(t)\rangle$. Here, we show these for $t\in[\tau_{S},13\tau_{S}]$, $\omega_{z}\approx \Omega(N)$, and $N=1,5$ to exemplify the quantities ${\mu_{\text{exp}}}(\langle \sigma_z \rangle)$ and ${\delta_{\text{exp}}}(\langle \sigma_z \rangle)$.}
\end{figure*}

Throughout our Letter, we estimate $D_{\text{eff}}$ numerically. $D_{\text{eff}}$ depends on $N$, $\Omega$, $\omega_{{z}}$, $\eta_1$, and $\rho(0)$. We approximate the latter by truncating the Hil{\-}bert space $\mathcal{H}$ to ${\mathcal{H}}_\text{trunc}$, choosing a phonon number cutoff $n_{{c}}$, such that ${\text{dim}}(\mathcal{H}_\text{trunc})=2(n_{{c}}+1)^N\lesssim2^{16}$ \cite{supplement}. For a given computational power  and increasing $N$, the description of the initial-state population by ${\text{tr}} \rho_\text{trunc}(0)$ becomes less representative, leading to increasing systematic uncertainties, illustrated in Fig.~\ref{fig1}(a). Here, the exponentially growing complexity becomes evident: ${\text{dim}}({\mathcal H}_\text{trunc})\approx 2^{22}$ is required to achieve ${\text{tr}} \rho_\text{trunc}(0)=0.99$ for $N=5$.
Figure 1(b) highlights the experimental controllability of $D_{\text{eff}}$.
 At $\{\Omega,\omega_{{z}}\}\approx\{2,1\}\times\omega_1$, the strong coupling to numerous modes leads to a maximum in $D_{\text{eff}}$.
For large $\omega_{{z}}$, the spin can get close to resonance with few modes only, the latter composing a comparatively small environment.
Further, the range of accessible values of $D_{\text{eff}}$ grows with increasing  $N$; see Fig.~\ref{fig1}(c).

We experimentally implement the single spin by two electronic hyperfine ground states of $^{25}$Mg$^+$  and add up to  four  $^{26}$Mg$^+$ to engineer the size of the bosonic environment spanned by $N$ (number of ions) longitudinal (axial)  motional modes. For details on our experimental setup, see Refs.~\cite{schaetz_towards_2007,clos_decoherence-assisted_2014}.
First, we prepare the spin state, $\rho_{{S}}(0) = |{\downarrow}\rangle \langle{\downarrow}|$, by optical pumping and initialize the  phonon state,  $\rho_{{E}}(0)$, by resolved sideband cooling~\cite{leibfried_quantum_2003} close to the ground state. In calibration measurements we determine that the modes are in thermal states with  $\bar{n}_{j=1\ldots N} \lesssim 1$, which effectively enhances $\eta_{j=1\ldots N}$. 
Next, we apply the spin-phonon interaction by continuously driving  Raman transitions with spin coupling rate $\Omega$ for variable duration $t$, where $\omega_{{z}}$ is the controllable detuning from  resonance~\cite{porras_mesoscopic_2008}.
Finally, we detect the spin by state-dependent fluorescence.
We choose to record $\langle \sigma_z(t) \rangle$, while we  numerically check that $\langle \sigma_{x,y}(t) \rangle$ feature similar behavior. To study dynamics for a large range of $D_{\text{eff}}$, we choose 95 parameter settings: We set $\omega_1/(2\pi)\approx 0.71\;$MHz
which corresponds to an effective spin-phonon coupling parameter $\eta_{1,\text{eff}}\equiv \eta_1\sqrt{2\bar{n}_1+1}\approx0.94$ for $\bar{n}_1=1$. For each $N=1\ldots 5$, we use a fixed $\Omega(N)/(2\pi)=\{0.73(1),0.95(3),1.28(3),1.37(3),1.58(5)\}\;$MHz and vary $\omega_{z}$ from $0$ up to $4\omega_1$~\cite{supplement}.

\begin{figure*}[t]
  \centering
  \includegraphics[]{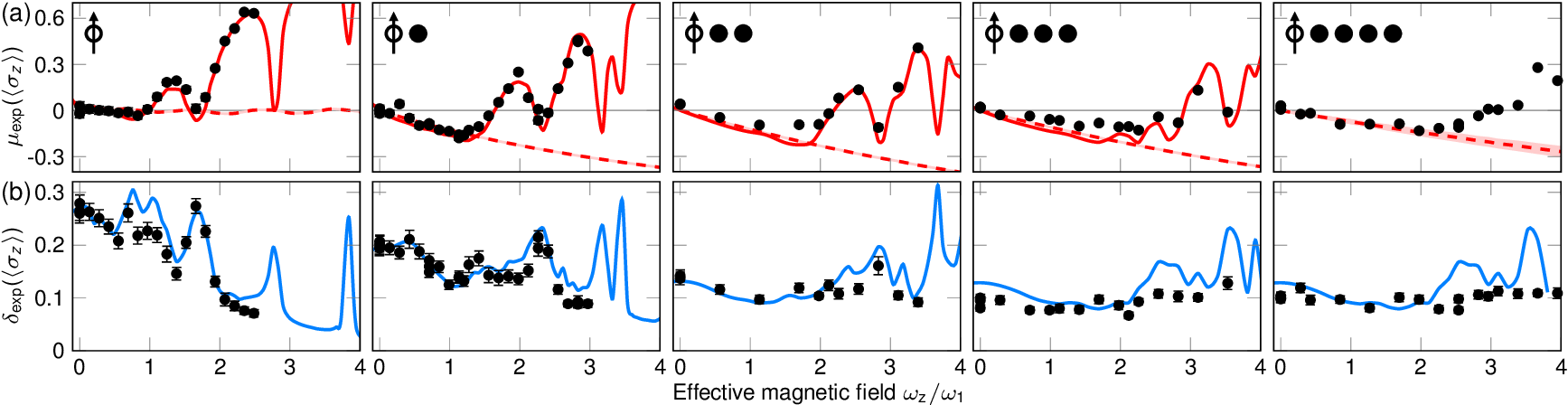}
   \caption{\label{fig3}Time averages and mean amplitudes of time fluctuations of $\langle \sigma_z(t) \rangle$. These are calculated from experimental traces (black dots, error bars: 1 s.d., derived from the underlying probability distribution of $\langle \sigma_z(t)\rangle$ \cite{supplement}) for varying $\omega_{{z}}$  and $N=1\ldots5$ and comparison to full ED (solid lines).  
(a) Increasing $\omega_{{z}}$ shifts the ground state of $H$, adjusts spin-mode couplings, and varies $D_\text{eff}$. Even for small systems, we find agreement of time averages with microcanonical averages, ${\mu_{\text{exp}}}(\langle \sigma_z \rangle)\approx{\mu_{\text{micro}}}(\langle \sigma_z \rangle)$ (dashed lines, shaded areas indicate systematic uncertainties). As $D_{\text{eff}}$ rapidly increases with $N$,  time averages follow microcanonical averages for a larger range of $\omega_{{z}}$. (b) ${\delta_{\text{exp}}}(\langle \sigma_{z} \rangle)$ gradually decreases with $N$ and correlated resonances in ${\mu_{\text{exp}}}(\langle \sigma_z \rangle)$ and  ${\delta_{\text{exp}}}(\langle \sigma_{z} \rangle)$ fade away, indicating that we tune our system from microscopic to mesoscopic size.}
\end{figure*}

In Fig.~\ref{fig2}, we present two sets of $\langle \sigma_z(t) \rangle$ for $N=1\ldots5$. 
Each data point is
 obtained by averaging over $r = 500$ repetitions yielding an expectation value  with  statistical uncertainty  $\propto 1/\sqrt{r}$.
We compare $\langle \sigma_z(t) \rangle$ with numerical full ED of Eq.~(\ref{fullham}) with ${\text{dim}}(\mathcal{H}_\text{trunc})\leq 2^{13}$.
As $N$ increases, the accuracy of numerical results decreases significantly. For $N=4$, we have ${\text{tr}} \rho_\text{trunc}(0)<0.72$. For $N=5$, since ${\text{tr}} \rho_\text{trunc}(0)<0.5$, we exclude numerical results in Figs.~\ref{fig2} and \ref{fig3}; here, even state-of-the-art full ED methods~\cite{beugeling_finite-size_2014} could consider ${\text{tr}} \rho_\text{trunc}(0)\lesssim0.75$ only \cite{supplement}.
For $\omega_{{z}}=0$ and $N=1$, we confirm oscillations of high and persisting amplitude due to the coupling to the only mode at $\omega_1$.
For increasing $N$, the spin couples to $N$ modes including  higher order processes, such that spin excitation gets distributed (entangled) into the growing bosonic environment.
Hence, coherent oscillations at incommensurate frequencies lose their common contrast and appear damped. After the transient duration $t /\tau_{{S}}\approx 1$,  with $\tau_{{S}}\equiv 2\pi/\Omega$, the spin observable remains close to its time average.
Still, the conservation of coherence of the evolution is evident in our measurements: Revivals of spin excitation due to the finite size of the system appear at $\tau_{\text{rev}}\sim 1/\overline{\delta E}$, where $\overline{\delta E}$ is the mean energy difference between modes.
And, for $\omega_{{z}}\approx\Omega$,  negative time averages of $\langle \sigma_z(t) \rangle$ indicate  equilibration of the system to the  ground state of $H$, biased by $\omega_{{z}}$. 
Both observations present strong independent evidence that our total spin-phonon system is near-perfectly isolated from external baths.
Independent measurements yield a decoherence rate of $\gamma_\text{dec}\tau_{S}\approx0.01$ \cite{supplement}.
This complements the agreement of experimental with numerical results, where we set $\gamma_\text{dec}=0$.

We analyze all recorded time evolutions, each containing $S\approx 100$  data points in the interval $[\tau_{{S}},13 \tau_{{S}}]$, by deriving
 time averages 
 \begin{equation}
   \label{eq:4}
   {\mu_{\text{exp}}}(\langle \sigma_z \rangle)\equiv \frac{1}{S}\sum_{t \in [\tau_{{S}},13 \tau_{{S}}]}\langle\sigma_z(t)\rangle
 \end{equation}
and mean amplitudes of time fluctuations 
\begin{equation}
  \label{eq:5}
{\delta_{\text{exp}}}(\langle \sigma_{z} \rangle)\equiv \sqrt{\frac{1}{S-1} \sum_{t \in [\tau_{{S}},13 \tau_{{S}}]} \left[\langle \sigma_z(t) \rangle - {\mu_{\text{exp}}}(\langle \sigma_z \rangle) \right]^2}.  
\end{equation}
The quantities are illustrated in two examples in Fig.~\ref{fig2}(c).
We plot these in Fig.~\ref{fig3} for $N=1\ldots5$ and  as a function of $\omega_{{z}}$, together with full ED results for $N=1\ldots 4$ (solid lines). Tuning $\omega_{z}$  across the maximum of $D_\text{eff}$, cf.~Fig.~\ref{fig1}(b), and comparing ${\mu_{\text{exp}}}(\langle \sigma_z \rangle)$ to numerically calculated microcanonical averages ${\mu_{\text{micro}}}( \langle\sigma_z \rangle)$ (dashed lines) \cite{supplement}, we find agreement for a larger range of $\omega_{{z}}$ when increasing $N$. This indicates an extended regime permitting thermalization. For large $\omega_{{z}}$, we observe its breakdown as the spin couples to an  environment of decreasing complexity.
 Finite-size effects are prominent in resonances of ${\mu_{\text{exp}}}(\langle \sigma_z \rangle)$ and ${\delta_{\text{exp}}}(\langle \sigma_{z} \rangle)$ for $N=1$, while their amplitudes gradually fade away for higher $N$.

\begin{figure}[htb]
  \centering
  \includegraphics[]{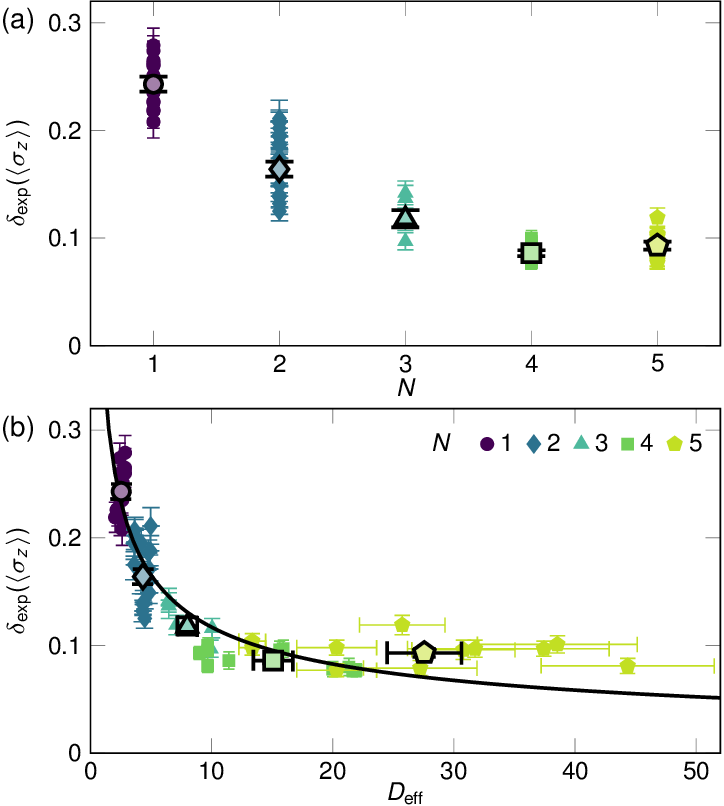}
  \caption{\label{fig4}Scaling of mean amplitudes of time fluctuations with $N$ and $D_\text{eff}$. (a) We plot ${\delta_{\text{exp}}}(\langle \sigma_{z} \rangle)$ (error bars: 1 s.d.) as a function of $N$. The spread for $N\leq 2$ highlights finite-size effects, and we show an average value for each $N$ (large symbols, error bars: 1 s.d.). For $N = 1\ldots 4$, we observe a decay that ceases for $N=5$. (b) ${\delta_{\text{exp}}}(\langle \sigma_{z} \rangle)$ as a function of calculated $D_\text{eff}$ (error bars: systematic uncertainties), which captures the dependence of the effective system size on all experimental parameters.
We compare to a scaling ${\delta_{\infty}}\propto 1/\sqrt{D_{\text{eff}}}$ (solid line), motivated for our system, and our measurements agree for $D_{\text{eff}}\lesssim25$. Further increasing $D_{\text{eff}}$, the system needs longer durations to resolve decreasing energy differences in the environment, unveiling the importance of time scales.}
\end{figure}

For further analysis, we postselect data points well described by microcanonical averages, i.e., with a deviation of less than $0.1$ \cite{supplement}. For those, we show the dependence of ${\delta_{\text{exp}}}(\langle \sigma_{z} \rangle)$ on $N$ in Fig.~\ref{fig4}(a). Although $N$ sets the size of the environment, the complexity of the spin-phonon coupling is tuned by $\Omega$, $\omega_z$, $\eta_1$, $\rho(0)$, and $N$, cf.~Figs.~\ref{fig1}(b) and \ref{fig1}(c). Consequently, we study the correlation between ${\delta_{\text{exp}}}(\langle \sigma_{z} \rangle)$ and $D_\text{eff}$ by combining our experimental results with numerical calculations of $D_\text{eff}$ in Fig.~\ref{fig4}(b). 
In general, mean amplitudes of time fluctuations are predicted to be upper bounded by $1/\sqrt{d_\text{eff}}$. For our system, we even find a proportionality, ${\delta_{\infty}}(\langle \sigma_{z} \rangle) \propto 1/\sqrt{D_{\text{eff}}}$:
We  motivate this scaling, illustrated by the solid line in Fig.~\ref{fig4}(b), by a heuristic derivation  considering pure initial states and infinite times, which relies on the eigenstate thermalization hypothesis~\cite{supplement}.
Our measurements feature such a scaling for $D_{\text{eff}}\lesssim25$, despite our nonidealized initial states and finite observation duration.
We observe that, for $D_\text{eff} \gtrsim 25$, measured mean amplitudes of time fluctuations do not further decrease. We attribute this to the fact that a system of increasing complexity features decreasing energy differences in its spectrum, corresponding to smaller relevant frequencies in the dynamics.
Explicitly, the system requires longer durations to approach theoretically predicted values.
Here, theory considers averages for infinite time, and does not make any prediction about relevant time scales in the dynamics.

In summary, we scale our trapped-ion system including its engineered environment up to relevant Hilbert space dimensions challenging state-of-the-art full ED.
We present time averages and fluctuations of a spin observable and exploit an effective dimension to study their dependence on the size of the system. We observe the emergence of quantum statistical mechanics within our isolated system despite its moderate size. Simultaneously, we monitor the coherent dynamics of thermalization, revealing the importance of initial and transient time scales by direct observation of the evolution towards thermal equilibrium. Thereby, we contribute to open questions in the field of thermalization~\cite{popescu_entanglement_2006,linden_quantum_2009,eisert_quantum_2015}.
 Our approach admits generating a multitude of initial conditions, choosing different system and  environment states, and preparing initial correlations~\cite{leibfried_quantum_2003,blatt_entangled_2008,kienzler_quantum_2015}. In addition, it allows us to measure a variety of observables \cite{leibfried_experimental_1996,leibfried_quantum_2003,kienzler_quantum_2015}. Applying those techniques, we can study, e.g., non-Markovianity of the dynamics, which is evidenced by revivals in the evolution, in detail~\cite{breuer_measure_2009,clos_quantification_2012}. Further, increasing the strength of the spin-phonon coupling, we can effectively expand the observable time span. Possible extensions include incorporating more and larger spins, tuning long-range interactions, adding external baths~\cite{porras_effective_2004,porras_mesoscopic_2008,myatt_decoherence_2000}, and propelling experimental quantum simulations.
Beyond numerical tractability, our experimental setup can be used as a test bed to assess the validity of approximated theoretical methods that address strong couplings to vibrational baths in a variety of fields, such as molecular and chemical physics.

Recently, we became aware of related studies with trapped ions, superconducting qubits, and ultracold atoms \cite{smith_many-body_2016,neill_ergodic_2016,kaufman_quantum_2016}.

\begin{acknowledgments}
We thank H.-P.~Breuer for discussions, M.~Enderlein, J.~Pacer, and J.~Harlos for assistance during the setup of the experiment, and M. Wittemer for comments on the manuscript. This work was supported by  the Deutsche Forschungsgemeinschaft [SCHA 973; 91b (INST 39/828-1 and 39/901-1 FUGG)], the People Programme (Marie Curie Actions) of the European Union's Seventh Framework Programme (FP7/2007-2013, REA Grant Agreement No: PCIG14-GA-2013-630955) (D.P.),
  and the Freiburg Institute for Advanced Studies (FRIAS) (T.S.).
\end{acknowledgments}


%


\clearpage
{\section{\large Supplemental Material}}

\section{Trapped-ion Hamiltonian}
Our system is described by the Hamiltonian presented in Eq.~(1), which we recast~\cite{porras_mesoscopic_2008} from:
\begin{equation}
H = \frac{\hbar \omega_z}{2} \sigma_z + \sum_{j = 1}^{N} \hbar \omega_j a^\dagger_j a_j + \frac{\hbar \Omega}{2} \left( \sigma^+ e^{i k_{\text L} z} + \sigma^- e^{-i k_{\text L} z}  \right).
\label{H.SB.app}
\end{equation}

As any two-level system, two electronic states can be interpreted as a pseudospin-1/2, represented by the Pauli operators
$\sigma_l\;(l=x,y,z)$. Controlling the effective energy difference of the states, $\hbar
\omega_z$, via the first term of the Hamiltonian is
equivalent to applying an effective magnetic field to the spin yielding a
Zeeman shift that lifts the degeneracy of the system in a controlled
way.

The second term describes the environment composed of $N$ harmonic oscillators, i.e., bosonic modes, with incommensurate frequencies $\omega_j$, and their excitations (phonons) constitute the environment of the spin in the otherwise closed quantum system. In Table~\ref{tab.explain} we list descriptions of all parameters in Eq.~\eqref{H.SB.app}. 
\renewcommand{\arraystretch}{1}
\begin{table*}[htb]
  \centering
  \footnotesize
  \begin{tabular}{p{0.4cm} p{2.8cm} p{3.5cm} p{6.0cm} p{3.7cm}}
    \hline\hline
    &parameter name&realization&physical interpretation&variation in experiment\\
    \hline
    $N$&number of axial motional modes&number of ions&number of axial harmonic oscillators spanning the bosonic environment&controlled by deterministic loading of ions, $N=1\ldots5$\\
    $\omega_j$&axial mode frequency&mutual Coulomb repulsion within axial trapping confinement&resonance frequencies of the harmonic oscillators (phonon energy $\hbar \omega_j$)&controlled by axial confinement, here fixed, $m=1\ldots N$\\
    $\Omega$&spin coupling rate (Rabi frequency)&Two-Photon Stimulated Raman transition (TPSR)&resonant coupling of the spin states $|{\downarrow}\rangle$ and \mbox{$|{\uparrow}\rangle$}, driving coherent oscillations between them&controlled by laser intensity, chosen for dedicated $N$ to maximize $D_{\text{eff}}$\\
    $\omega_z$&effective magnetic field&TPSR detuning from resonant (i.e., carrier) transition&lifts degeneracy of coupled spin states $|{\downarrow}\rangle$ and $|{\uparrow}\rangle$, i.e., introduces a bias and alters the total ground state&controlled by variable detuning\\
    $\eta_j$&spin-phonon coupling (Lamb-Dicke) parameters&momentum kick by the two photons of the TPSR&momentum conservation when creating and annihilating phonons: $\eta^2=$ (photon recoil energy)/(phonon energy); for $\eta\ll 1$: weak coupling (only carrier transition, i.e., spin flip)&controlled via $\omega_1$ (confinement), here fixed, $\eta_1=0.54$\\
    \hline\hline
  \end{tabular}
  \caption{\label{tab.explain}Detailed description of all parameters in the Hamiltonian.}
\end{table*}

We couple spin and  phonons through the photon recoil term $e^{i
k_{\text L} z}$.
Here, $z$ is the position of the ion, carrying the spin, and
$k_{\text L}$ represents the effective wave vector of the optical
field, driving a two-photon stimulated Raman transition with spin coupling rate $\Omega$, the so-called Rabi frequency (see below).
The related recoil, $\hbar k_{\text L}$, is required for creating
(annihilating) phonons within the modes while flipping the spin,
that is, assures energy and momentum conservation of the spin-phonon
coupling.
We want to emphasize that, in principle, the optical drive induces
coherent transitions only, and, consequently, does not represent
a channel to an external bath.
The last term in Eq.~\eqref{H.SB.app} incorporates  both, the
term $(\hbar \Omega/2) \sigma_x$, which does not affect the motional
state, and the spin-phonon coupling $(\hbar \Omega/2)(\sigma^+ C + \sigma^- C^\dagger)$, where the spin-flip operators can be related via $\sigma_x = (\sigma^+ + \sigma^-)/2$.

The position $z$ is expressed as 
\begin{equation}
  \label{eq:1}
k_{\text L} z =
\sum_{j = 1}^N \eta_j \left(a^\dagger_j + a_j \right),  
\end{equation}
with corresponding spin-phonon coupling (Lamb-Dicke) parameters
\begin{equation}
  \label{eq:2}
\eta_j = {\mathcal M}_j k_{\text L} \sqrt{\frac{\hbar}{2 M \omega_j}}.
\end{equation}
Here, $M$ denotes the mass of the ion chain and ${\mathcal M}_j$
is the wavefunction amplitude of mode $j$.
Further, it is convenient to use $\eta_j =\mathcal{M}_j \sqrt{\omega_1/\omega_j}
\eta_1$, with $\eta_1$ describing the longitudinal center-of-mass
mode ($j = 1$).%

For values $\eta_j \ll 1$, we find linear/weak spin-phonon coupling
that is commonly explored in trapped-ion experiments and 
\begin{equation}
  \label{eq:3}
C \simeq i \sum_{j=1}^N \eta_j (a_j^\dagger + a_j).
\end{equation}
We, however, tune  $\eta_j\simeq 1$ to exploit higher-order phonon emission/absorption terms and in our experiments the full expression,
\begin{equation}
  \label{eq:4}
C = {\text{exp}}\left[{i \sum_{j=1}^N \eta_j (a_j^\dagger + a_j)}\right]-1,  
\end{equation}
is relevant.
Driving at $\Omega \simeq \omega_j$ has two main consequences: 
First, the coupling of the spin states and the spin-phonon coupling get further increased and second, the
individual transitions between single and multiple modes cannot get resolved anymore.
Note, that the coupling strength can be effectively enhanced, as $\eta_j$ is proportional to $\mathcal{M}_j$ and, thus, scales with the Fock state $|n_j\rangle$: 
\begin{equation}
  \label{eq:5}
\eta_j' \approx \eta_j \sqrt{1 + 2 n_j}.  
\end{equation}

In our work, we explicitly make use of the nonlinear/strong coupling and bring our system into a nonintegrable regime.
The system Hamiltonian is exactly solvable in the following cases: 
(i) $N = 1$ and $\omega_z = 0$, in which it corresponds to the quantum Rabi model~\cite{braak_integrability_2011}, and (ii) $\Omega = 0$, in which spins and phonons are not coupled at all. 
In any other parameter regime, it can be considered a nonintegrable Hamiltonian, by the applicable definitions of
nonintegrability~\cite{caux_remarks_2011}: (i) it does not have an exact analytical solution, and (ii) there is not a number of independently conserved observables that equals the number of degrees of freedom in the system. 

The full quantum dynamics of the thermalization process may be treated with quasi-exact methods, since relevant aspects of thermalization, such as time scales and size of fluctuations, critically depend on the set of frequencies involved in the dynamics. 
In turn, standard approaches for the spin-boson model, such as path-integral methods or generalized master Eq.~\cite{weiss_quantum_1999}, rely on the assumption that the phonon bath is characterized by a continuous spectrum of frequencies, and thus, they cannot be applied to our mesoscopic system. 
We access the dynamics by full ED, which provides all eigenstates and eigenenergies of the total (truncated) Hamiltonian, and has been done in recent theoretical studies that investigate similar physics in other systems~\cite{rigol_thermalization_2008, rigol_breakdown_2009,santos_onset_2010,beugeling_finite-size_2014,kim_testing_2014,cazalilla_focus_2010}.

\section{Inverse Participation Ratio}
The Inverse Participation Ratio ({\text{IPR}}) is used as an ergodicity measure to quantify the ability of a system to thermalize~\cite{srednicki_approach_1999,linden_quantum_2009}.
Here, thermalization describes a process in a closed quantum system, where values of observables (of any subsystem) equal microcanonical averages~\cite{deutsch_quantum_1991,srednicki_chaos_1994}, while the total system remains in an entangled state during the unitary evolution. In general, it requires for (any) few-body observable: 
(i) the equivalence between time averages and microcanonical averages, and 
(ii) decreasing mean amplitudes of time fluctuations around a mean value as a function of the system size. 

\begin{figure}[htb]
  \centering
  \includegraphics[]{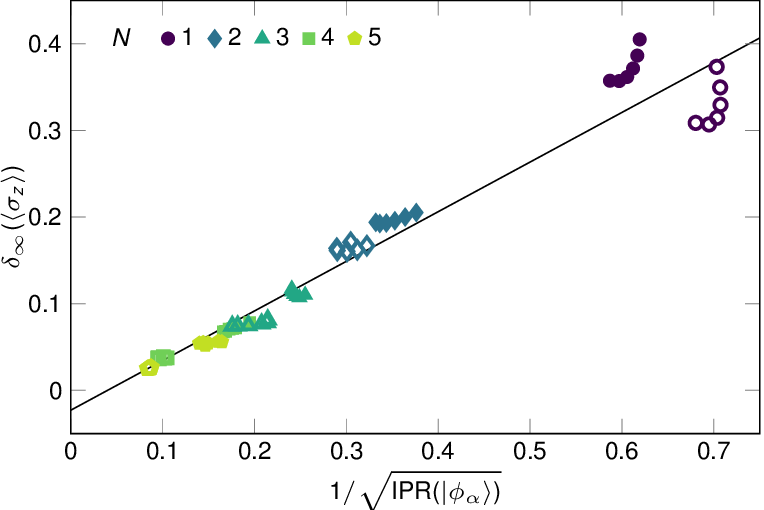}
   \caption{\label{fluctVsIPRsz}Numerical study of the scaling of mean amplitudes of time fluctuations. Filled and empty symbols correspond to initial states $|\phi_\alpha\rangle$ with one and two phonons per mode, respectively. Time fluctuations averaged over infinite time are calculated using Eq.~\eqref{time.fluctuations}. Parameters:
$N = 1$, $n_c =$ 20, $\Omega/(2\pi)= $ 0.7 MHz. 
$N = 2$, $n_c =$ 10, $\Omega/(2\pi)=$ 1.0 MHz. 
$N = 3$, $n_c =$ 6, $\Omega/(2\pi)=$ 1.3 MHz.
$N = 4$, $n_c = $ 5, $\Omega/(2\pi)=$ 1.4 MHz.
$N = 5$, $n_c = $ 4, $\Omega/(2\pi)=$ 1.6 MHz. For all values of $N$, we set $\omega_1/(2\pi)=0.7\,$MHz and take $\omega_z =$ $\Omega/4, \dots, \Omega/2$, in steps of $\Omega/20$. This numerical study substantiates the validity of our heuristic derivation of Eq.~\eqref{first.final.relation} (solid line), and our experimental observation of the scaling with effective dimension.}
 \end{figure}

The IPR is defined for a set of energy eigenstates
$\{| \psi_\beta \rangle\}$
and an initial pure state
$| \phi_\alpha \rangle$ 
as 
\begin{equation}
  \label{eq:6}
{\text{IPR}}(| {\phi_\alpha} \rangle) = \frac{1}{\sum_{\beta} |c_\beta(\alpha)|^4 }.  
\end{equation}
Recent numerical work shows that systems with increasing values of {\text{IPR}} fulfill condition (i) more accurately, see for example Ref.~\cite{beugeling_finite-size_2014}, where it is shown that deviations of diagonal ensemble predictions from microcanonical ensemble averages decrease with IPR.
Intuitively, we may expect that it also leads to condition (ii): We argue that a large IPR involves a large set of eigenfrequencies in the dynamics of few-body observables and destructive interferences can lead to a suppression of the size of fluctuations on average. However, large time fluctuations can still appear, but during ever shorter durations. Indeed, an upper bound on these fluctuations can be obtained rigorously based on Ref.~\cite{linden_quantum_2009}. 
For our system, one can find a heuristic derivation that mean amplitudes of time fluctuations monotonically decrease as a function of IPR (see below). The derivation relies on the Eigenstate Thermalization Hypothesis (ETH), plausible assumptions on the many-body eigenstates, and a large number of participating states (parameter regimes with large IPR). 
We mathematically formulate the ETH and explain it in detail in the last section of this Supplemental Material. To examine the validity of our derivation, we perform numerical calculations and compare them to the derived scaling: 
\begin{equation}
{\delta_{\infty}}^2(\langle \sigma_z \rangle) \propto 
\frac{1}{{\text{IPR}}(|\phi_\alpha\rangle)},
\label{first.final.relation}
\end{equation}
which is valid for initial pure states and infinite times. In Fig.~\ref{fluctVsIPRsz} we show numerical results for ${\delta_{\infty}}^2(\langle \sigma_z \rangle)$ for a variety of initial conditions and system parameters as a function of IPR, and the estimate based on Eq.~\eqref{first.final.relation} is shown as a solid line. We find reasonable agreement even for small values of IPR.

In our study, the initial state is a product state of a  pure spin state and a motional state cooled close to its ground state, i.e., a mixed phonon state, yet with low average occupation. Thus, we need to find a method to average the IPR and get an appropriate measure of the effective dimension of an initial mixed state. In Ref.~\cite{short_quantum_2012} it is shown that, considering a mixed initial state $\rho(0) = \sum_{\alpha} w_\alpha |\phi_\alpha \rangle \langle \phi_\alpha|$, an upper bound to mean amplitudes of time fluctuations around time-averaged states
can be found from the definition 
${\text{IPR}}^{-1}(\rho(0)) = \sum_\beta p_\beta^2$, with 
$p_\beta = \sum_\alpha w_\alpha |c_\beta(\alpha)|^2 $. 
However, this approach has the disadvantage that the IPR takes large values for highly mixed initial states, even in the case of uncoupled systems, since then $c_\beta(\alpha) = \delta_{\beta,\alpha}$ and ${\text{IPR}}^{-1}(\rho(0)) = \sum_\alpha w_\alpha^2$. 
We choose to define an effective dimension by averaging over the {\text{IPR}}s of each state in the mixture, which directly leads to our definition in Eq.~(5). This definition yields $D_{\text{eff}}  = 1$ for uncoupled  Hamiltonians. It takes large values only if interactions lead to a large number of  eigenstates participating in the dynamics, allowing us to identify regions where the system should be able to thermalize. 

The effective dimension $D_{\text{eff}}$ is a weighted measure of the  number of coupled basis states that are required to express the uncoupled basis states. For example, consider a pure state $|\phi_\alpha\rangle$ in the uncoupled basis,
represented in the coupled basis $\{|\psi_\beta\rangle\}$, i.e., $|\phi_\alpha\rangle = \sum_\beta c_\beta (\alpha) |\psi_\beta\rangle$.
Typically, most coefficients $|c_\beta (\alpha)|^2\ll 1$, such that a significant value of $D_{\text{eff}}$ is obtained only after considering a large number of states $|\psi_\beta\rangle$, each of them contributing with a small fraction to
the final effective dimension. In addition to this, part of our our initial state is a thermal state that incorporates  a large number of uncoupled states, $|\phi_\alpha\rangle$. Our definition is chosen such that we do not artificially increase $D_{\text{eff}}$ by simply adding these states, but we renormalize $D_{\text{eff}}$ according to their contribution to the mixed state.

\section{Numerical diagonalizations and their systematic uncertainties}

We perform full ED of Eq.~(1) to calculate $\langle \sigma_{z} (t) \rangle$ and corresponding {micro\-canonical} averages ${\mu_{\text{micro}}}(\langle \sigma_{z} \rangle )$ up to $n_{\text c}$. 
For the calculations, we use Matlab on a workstation with a quad-core 3.7 GHz processor
and 64 GB RAM. Besides limitations in computation time, a fundamental issue when calculating the
full eigensystem of a $d \times d$ matrix lies in the size of the available main memory that is required to store and process this matrix. Our computer is unable to fulfill this task (already) for systems of $N =5$ ions with a cutoff at $n_c = 6$ phonons per mode that corresponds to a matrix of dimension $d^2\approx2^{15} \times 2^{15}$.
For our experimental parameters, this forces us to neglect more than half of the initial-state population, ${\text{tr}}\rho_\text{trunc}(0)< 0.46$, while $d\approx 1.5 \times 2^{17}$ would be required for ${\text{tr}}\rho_\text{trunc}(0)= 0.75$ and $d\approx 1.4\times 2^{25}$ for ${\text{tr}}\rho_\text{trunc}(0)= 0.99$.
 As a benchmark of our numerical effort, we compare this to recent theoretical studies that diagonalize maximum sizes of: $d \approx 6.6 \times 10^4 \approx 1.0 \times 2^{16}$~\cite{luitz_many-body_2015}, $d \approx 5.2 \times 10^4\approx1.6\times 2^{15}$~\cite{mondaini_eigenstate_2016}, and $d \approx 1.2 \times 10^5\approx1.8\times 2^{16}$~\cite{beugeling_finite-size_2014}, while the authors state that those are the maximum sizes tractable by their methods.
In Table~\ref{tab.numerics} we list parameters used for all presented numerical results.

For calculations of $D_{\text{eff}}$, we use an approximation, i.e., a block-diagonalization procedure that exploits the band structure of our Hamiltonian matrix. We express Eq.~(1) in the basis of eigenstates of the uncoupled system 
$|\phi_\alpha \rangle = |s\rangle |n_1\rangle_1 \dots |n_N\rangle_N$, with $s = \{\uparrow, \downarrow\}$, and $|n_j \rangle$ the Fock state with $n$ phonons in the $m$th motional mode. The Hamiltonian takes the band matrix form,
\begin{equation} 
H_{\alpha_1,\alpha_2} = \langle \phi_{\alpha_1}| H | \phi_{\alpha_2} \rangle 
= E^{(0)}_{\alpha_1} \delta_{\alpha_1,\alpha_2} + V_{\alpha_1,\alpha_2},
\end{equation} 
with $V_{\alpha_1,\alpha_2}$ the matrix of nondiagonal elements. 
We order the states according to their energy such that the diagonal entries $E^{(0)}_\alpha$ grow with index number $\alpha = 1 \ldots {\text{dim}}(\mathcal{H}_\text{trunc})$. 
The matrix $V_{\alpha_1,\alpha_2}$ couples only those states that fulfill 
$|E^{(0)}_{\alpha_1} - E^{(0)}_{\alpha_2}| \lesssim  |V_{\alpha_1,\alpha_2}|$.
Further, we express the initial state $\rho(t = 0) = \sum_\gamma w_\gamma |\phi_\gamma \rangle \langle \phi_\gamma |$, calculate $D_{\text{eff}}$ of each of the initial states $| \phi_\gamma \rangle$ by diagonalizing $H^{[\gamma]}_{\alpha_1,\alpha_2}$, and keep only quantum states $\alpha_1$ and $\alpha_2$ close to $\gamma$ such that ($|\alpha_1 - \gamma|, |\alpha_2 - \gamma|) < N_{\text{states}}/2$, where $N_{\text{states}}$ denotes the number of states kept to calculate the {\text{IPR}} of each of the states $| \phi_\gamma \rangle$. Thus, $H^{[\gamma]}_{\alpha,\beta}$ is projected onto a subspace of 
$N_{\text{states}} < {\text{dim}}({\mathcal H}_\text{trunc})$, that optimally contributes to $D_{\text{eff}}$ of $| \phi_\gamma \rangle$. Finally, we diagonalize $H^{[\gamma]}_{\alpha_1,\alpha_2}$, calculate the contribution of $| \phi_\gamma \rangle$, and increase $N_{\text{states}}$ by steps of $1000$, until the resulting values of $D_{\text{eff}}$ vary by less than $1\%$ $(5\%$ for $N=5)$. 
For example, for $N=4$, we use $n_{\text c} = 12$, corresponding to ${\text{dim}}(\mathcal{H}_\text{trunc})\approx 2^{16}$, which converges using  $N_{\text{states}} = 5000$. However, in the case of $N = 5$ and $n_{\text{c}} = 7$, we are (again) at the limit of our computational power, because we need $N_{\text{states}} = 20000$ for converged values of $D_{\text{eff}}$.

\begin{figure}[htb]
  \centering
  \includegraphics[]{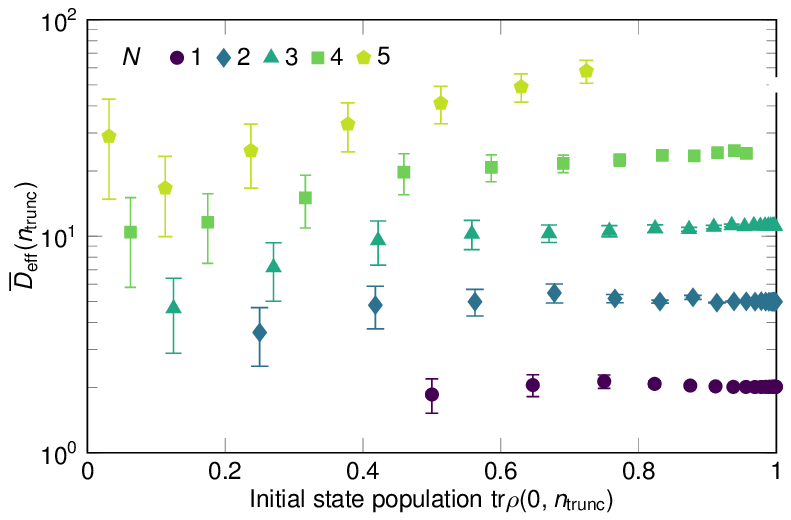}
   \caption{\label{supp_uncertainties}Numerical study of $D_{\text{eff}}$ as a function of truncated initial-state population. We plot $\overline{D}_{\text{eff}}(n_{\text{trunc}})$ for $N=1\ldots5$ (details see text). The truncation $n_{\text{trunc}} = 1$ corresponds to lowest ${\text{tr}} \rho_\text{trunc}(0)$, while $n_{\text{trunc}}$ is increased in steps of one up to its maximum value $n_{\text c}$ (limited by our computer memory; see Tab.~\ref{tab.numerics}). The parameters are the same as in Fig.~1(c): $\omega_1/(2\pi)= 0.71$ MHz, $\eta_1=0.54$, $\bar{n}_{j=1\ldots N}=1$, $\Omega/(2\pi)= \{0.71, 0.97, 1.21, 1.46, 1.68\}$ MHz, $\omega_{\text z} = 0.8\omega_1$. For $N = 1\ldots4$ and for ${\text{tr}} \rho_\text{trunc}(0) > 0.5$, we find that $\overline{D}_{\text{eff}}(n_{\text{trunc}})$ agree within their attributed systematic uncertainties (error bars) with numerically most accurate values $D_{\text{eff}} (n_{\text c})$ which have ${\text{tr}} \rho(0, n_{\text c}) > 0.96$. This suggests that our estimation of $D_{\text{eff}}$ yields reasonable results, yet with large systematic uncertainties, even if about $0.5$ of the initial-state population are neglected in the numerical calculations.}
 \end{figure}
\renewcommand{\arraystretch}{1}
\begin{table*}[htb]
  \centering
  \footnotesize
  \begin{tabular}{lc|cccccccccc}
    \hline\hline
    &$N$
    &\multicolumn{2}{c}{$1$}
    &\multicolumn{2}{c}{$2$}
    &\multicolumn{2}{c}{$3$}
    &\multicolumn{2}{c}{$4$}
    &\multicolumn{2}{c}{$5$}\\\cline{1-2}
    $\text{Fig.}$&
    &$n_{\text c}$ &${\text{tr}}\rho_\text{trunc}(0)$
    &$n_{\text c}$ &${\text{tr}}\rho_\text{trunc}(0)$
    &$n_{\text c}$ &${\text{tr}}\rho_\text{trunc}(0)$
    &$n_{\text c}$ &${\text{tr}}\rho_\text{trunc}(0)$
    &$n_{\text c}$ &${\text{tr}}\rho_\text{trunc}(0)$\\
    \hline
    $\text{1(b)}$ &$D_{\text{eff}}$       &  &    &  &    &$10$&$0.94$&  &    & &\\
    $\text{1(c)}$ &$D_{\text{eff}}$       &$20$&$1.00$&$20$&$1.00$&$20$&$1.00$&$12$&$0.96$&$7$&$0.72$\\
    $\text{2}$ &$\langle \sigma_z(t) \rangle     $ &$20$&$1.00$&$20$&$1.00$&$11$&$0.97$&$ 6$&$0.72$&$(4 $&$0.35)$\\
    $\text{3(a)}$ &${\mu_{\text{micro}}}(\langle \sigma_z \rangle) $&$20$&$1.00$&$20$&$1.00$&$16$&$0.99$&$ 8$&$0.86$&$5$&$0.46$\\
    $\text{3(a)}$ &${\mu_{\text{exp}}}(\langle \sigma_z \rangle) $ &$20$&$1.00$&$20$&$1.00$&$ 9$&$0.93$& $5$&$0.62$&$(3 $&$0.23)$\\
    $\text{3(b)}$ &${\delta_{\text{exp}}}(\langle \sigma_{z} \rangle)$   &$20$&$1.00$&$20$&$1.00$&$ 9$&$0.93$&$ 5$&$0.62$&$(3 $&$0.23)$\\
    $\text{4}$ &$D_{\text{eff}}$       &$20$&$1.00$&$20$&$1.00$&$20$&$1.00$&$12$&$0.96$&$7$&$0.64$\\
    \hline\hline
  \end{tabular}
  \caption{\label{tab.numerics}Parameters used for numerical calculations. Cutoff in Fock space $n_{\text c}$ and truncated initial-state population ${\text{tr}}\rho_\text{trunc}(0)$ used in the numerical calculation for the different Figures.
}
\end{table*}

In the following, we describe our procedure to yield final values of $D_{\text{eff}}$ including a measure of systematic uncertainties accounting for population which has been neglected by the truncation, ${\text{tr}} \rho(0)<1$. We evaluate $D_{\text{eff}}$ as a function of truncation $0 < n_{\text{trunc}} \leq n_{\text c}$ for all parameter settings presented in our manuscript (for a list of $n_{\text c}$ see Tab.~\ref{tab.numerics}). For each parameter setting, we calculate the truncated initial-state population ${\text{tr}} \rho(0,n_{\text{trunc}} )$, take $D_{\text{eff}} (n_{\text{trunc}})$ as a first bound $D_{\text{eff}, 1}$, linearly extrapolate from it a second bound $D_{\text{eff}, 2}$ with the slope extracted from $D_{\text{eff}}(n_{\text{trunc}}-1)$, and plot $\overline{D}_{\text{eff}} \equiv (D_{\text{eff}, 1}+D_{\text{eff}, 2})/2$ as a function of ${\text{tr}} \rho(0,n_{\text{trunc}} )$. In Fig.~\ref{supp_uncertainties}, we  show an example of this. The error bars represent our attributed systematic uncertainties, given by $\pm\left |D_{\text{eff}, 1}-D_{\text{eff}, 2}\right |/4$. For $N=1\ldots4$ and all parameter settings, we find that, for ${\text{tr}} \rho(0) > 0.5$, all extracted $\overline{D}_{\text{eff}}$ agree within their uncertainties with corresponding $D_{\text{eff}} (n_{\text c})$, which, in turn, we consider most accurate, because for $N\leq4$, it is ensured that ${\text{tr}} \rho(0, n_{\text c}) > 0.96$. This suggests that our procedure yields reasonable values for $D_{\text{eff}}$, yet with large uncertainties, even when up to $0.5$ of the initial-state population is neglected in the numerical calculation, as it is the case for $N = 5$. We use for final values $D_{\text{eff}} \equiv \overline{D}_{\text{eff}}(n_{\text c})$, throughout the manuscript. 

To calculate the microcanonical average, ${\mu_{\text{micro}}}(\langle \sigma_{z} \rangle)$, we calculate mean and standard deviation of the energy with respect to the initial state, $\overline{E}=\langle H \rangle$, and 
\begin{equation}
  \label{eq:7}
\Delta E = \sqrt{\langle H^2 \rangle - \overline{E}^2}.  
\end{equation}
We define a microcanonical ensemble by using a Gaussian distribution of the form:
\begin{equation}
{\mu_\text{micro}}(\langle \sigma_{z} \rangle) = \sum_\beta P_\beta\;(\sigma_z)_{\beta,\beta},
\end{equation}
where
\begin{equation}
P_\beta = \text{exp}\left[-\frac{(E_\beta-\overline{E})^2}{(\Delta E/2)^2}\right]/ \sum_\beta \text{exp}\left[-\frac{(E_\beta-\overline{E})^2}{(\Delta E/2)^2}\right].  
\end{equation}

The ETH implies that $(\sigma_z)_{\beta,\beta}$ is a smooth function of the energy $E_\beta$, and, thus, the microcanonical prediction should not depend on the details of the energy distribution.
Further, we apply a procedure similar to the one described above for $D_{\text{eff}}$, but based on ED (no block-diagonalization), to provide final values of ${\mu_{\text{micro}}}(\langle \sigma_{z} \rangle)$ including attributed systematic uncertainties. We include numerical calculations of microcanonical averages in Fig.~3 and Fig.~\ref{supp_deff_deth}, even for $N=5$, where $\text{tr}\rho_\text{trunc}<0.46$.
This estimation is only used to detect regions in
parameter space where the measured mean value is close to the microcanonical average (cp.~Fig.~\ref{supp_deff_deth}), and is not used in our further analysis.
Additionally, numerical calculations show a faster convergence of
${\mu_{\text{micro}}}(\langle \sigma_{z} \rangle)$ with respect to the cutoff $n_c$, since that average is less sensitive to the total number of quantum many-body states than IPR and $D_\text{eff}$.

\section{Experimental setup}
\begin{figure*}[htb]
  \centering
  \includegraphics[]{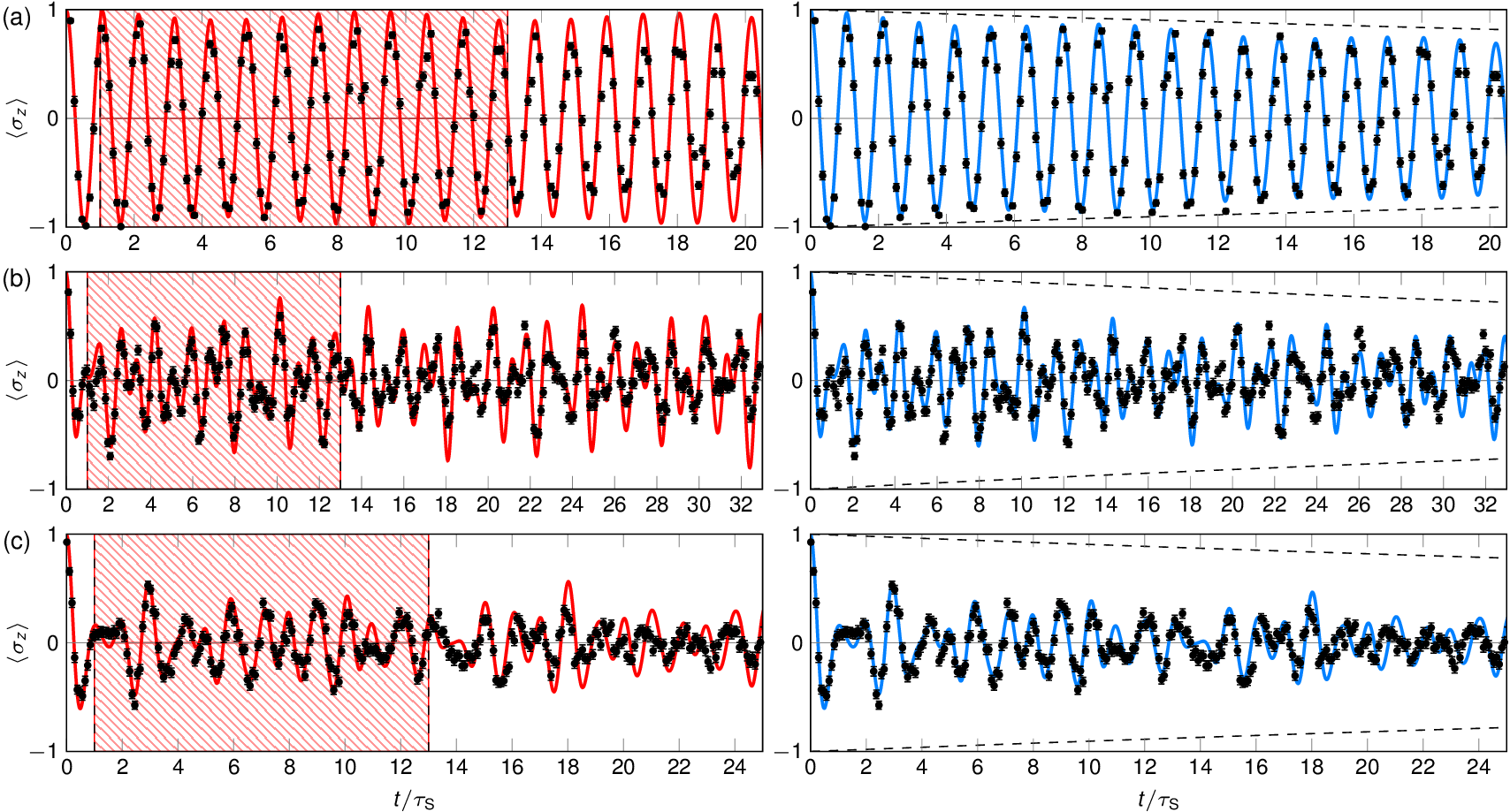}
   \caption{\label{supp_dec_flops}Test for unitarity of the experimentally measured time evolutions. (left) Measured values with statistical uncertainties (black dots) and full ED of the unitary time evolution, i.e., with no decoherence (solid line). The shaded area emphasizes the time span that is considered to derive ${\mu_{\text{exp}}}(\langle \sigma_z \rangle)$ and ${\delta_{\text{exp}}}(\langle \sigma_{z}\rangle)$ (see Figs.~2 and 3), $1\leq t/\tau_{\text S} \leq 13$. For $t/\tau_{\text S}\lesssim15$, we find our experimental data in good agreement with the numerical calculations. (right) The remaining discrepancy can be captured by implying a weak decoherence indicated by the dashed line $\langle\sigma_{z}'(t)\rangle=\langle \sigma_z(t) \rangle e^{-\gamma_\text{dec} t}, \;\gamma_\text{dec}\tau_{\text S}=0.01$, which is in agreement with independent calibration measurements. Parameters for the experiments:
(a) $N=1$, $\{\omega_1,\Omega,\omega_{\text z}\}/(2\pi)= \{1.92(1), 0.685(5), 0.00(1)\}$ MHz, $\bar{n}_1=0.05(3)$
(b) $N=1, \{\omega_1,\Omega,\omega_{\text z}\}/(2\pi)= \{0.73(1), 0.828(5), 0.00(1)\}$ MHz, $\bar{n}_1=0.6(1)$
(c) $N=2, \{\omega_1,\Omega,\omega_{\text z}\}/(2\pi)= \{0.71(1), 0.940(5), 0.00(1)\}$ MHz, $\bar{n}_{\{1,2\}}=\{0.4(1),0.6(1)\}$.
This comparison further justifies to describe our system by means of unitary time evolution, and, consequently, treat it as fully isolated within our experimental observation.
}
\end{figure*}

We employ a linear radio-frequency (rf) Paul trap with a drive frequency $\Omega_{\text{rf}}/(2\pi)\approx 56$ MHz to trap Mg$^+$ isotopes with secular frequencies of $\omega_{\text{x,y}}/(2\pi)\approx \{4.0,4.6\}$ MHz (radial direction) and
 $\omega_1/(2\pi)\approx 0.7$ MHz (axial direction)~\cite{schaetz_towards_2007}.
Using a photoionization laser (wavelength $\lambda \approx 285$ nm), we isotope-selectively load $1\times^{25}$Mg$^+$ and $(N-1)\times^{26}$Mg$^+$, and prepare identical spatial configurations in all measurements~\cite{hume_two-species_2010}. Two electronic ground states of the hyperfine manifold of $^{25}$Mg$^+$ (nuclear spin $I_{\text{25Mg}}=5/2$) constitute the pseudospin, $|{\downarrow}\rangle \equiv 3S_{1/2}|F{=}3,m_F{=}3\rangle$ and $|{\uparrow}\rangle \equiv 3S_{1/2}|F{=}2,m_F{=}2\rangle$, where $F$ and $m_F$ denote the total angular momentum quantum numbers of the valence electron. Note, that $^{26}$Mg$^+$ has no nuclear spin ($I_{\text{26Mg}}=0$) and consequently, in our realization, it constitutes (only) to the motional mode structure.

To address transitions from $3S_{1/2}$ to either $3P_{1/2}$ or $3P_{3/2}$ in $^{25}$Mg$^+$ at wavelengths around $280$ nm, we use dedicated all solid state laser systems~\cite{clos_decoherence-assisted_2014,friedenauer_high_2006}: A laser beam with $\lambda_{\text{BD}} \approx 279.64$~nm (labeled BD) is tuned $\Gamma_{\text{nat}}/2$ below the $3S_{1/2}$ to $3P_{3/2}$ transition with natural line width $\Gamma_{\text{P3/2}}/(2\pi) = 41.8(4)$~MHz~\cite{ansbacher_precision_1989} and aligned with a magnetic quantization field $\vert \vec{B} \vert \simeq0.58$ mT. The $\sigma^+$-{po\-la\-rized} light with intensity $I_{\text{BD}}\approx 0.5 I_{\text{sat}}$ (saturation intensity $I_{\text{sat}}\simeq 2500$ W/m$^2$) is used for Doppler cooling to $\simeq 1\;\text{mK}$ and optical pumping to $|{\downarrow}\rangle$. This state provides efficient, state sensitive detection via the closed cycling transition  to $3P_{3/2}$~$|{4,4}\rangle$ to discriminate between the ground state manifolds. When BD is shifted near resonance (detuning $\leq 2\pi \times 4$~MHz), we record photon-count histograms with averages corresponding to count rates of $100$ ms$^{-1}$ for all $F{=}3$ states and $2$ ms$^{-1}$ for the $F{=}2$ states.
We can detect experiments in which the ion-chain order is disturbed by a systematic change in histogram distributions, and postselect experiments where the desired ion-order is achieved, which is true in $\{97.1(3),82(2),81(1),79(2)\}\;\%$ of all our experiments for $N=2\ldots 5$.
 Two additional laser beams with $\lambda_{\text{RP}} \approx 280.35$ nm (labeled RP1 and RP2) are superimposed to BD and used for repumping electronic state population from $3S_{1/2}|F{=}3,m_F{<}3\rangle$ (RP1) and $3S_{1/2}|F{=}2\rangle$ (RP2) to $|{\downarrow}\rangle$ via $3P_{1/2}$ (with $\Gamma_{\text{P1/2}}/(2\pi) = 41.3(3)$ MHz ~\cite{ansbacher_precision_1989}). 
Further, two perpendicular laser beams at $\lambda_{\text{Raman}} \approx 279.61$~nm (labeled RR and BR) with a $k$-vector difference, $k_L$ along the axial direction of the trap (ion-chain alignment) enable ground-state cooling of axial motional modes. This is achieved via sideband cooling with two-photon stimulated Raman transitions (TPSR) detuned by $\Delta/(2\pi)= + 100(10)$ GHz from the $3S_{1/2}$ to $3P_{3/2}$ transition. In the regime of resolved sidebands, e.g., for $\omega_1/(2\pi)= 1.90(1)$ MHz, $N\leq 3$, and $\Omega/(2\pi)\leq 0.5$ MHz, we routinely achieve occupation numbers $\bar{n}_{j=1\ldots N}<0.1$  by an iterative pulsed sideband cooling procedure.
For  $\omega_1/(2\pi)= 0.71(1)$ MHz, we use $\Omega/(2\pi) \approx 0.25$ MHz and achieve $\bar{n}_{j=1\ldots N}<1$ for $N\leq5$ by addressing several sideband transitions simultaneously. In Table~\ref{tab.params}, we summarize relevant experimental parameters that are determined with dedicated calibration measurements, e.g., mode-temperature measurements~\cite{wineland_experimental_1998}.
\begin{table}[hb]
  \centering
\begin{tabular}{lccccc}
    \hline\hline
     $N$&$1$  &$2$  &$3$  &$4$  &$5$\\
    \hline
     $\omega_1/(2\pi \text{MHz)}$ &$ 0.724(2) $&$ 0.707(2) $&$ 0.707(2) $& $0.708(2) $&$ 0.709(2) $\\
     $\omega_N/\omega_1$ &$ 1.00(1) $&$ 1.73(1) $&$ 2.41(1) $&$ 3.05(1) $&$ 3.67(1) $\\
     $\omega_{\text z}/(2\pi\text{MHz)} $&$[0, 1.8] $&$[0, 2.2] $&$[0, 2.4] $&$[0, 2.5] $&$[0, 2.8] $\\
     $\Omega/(2\pi\text{MHz)} $&$ 0.73(1) $&$ 0.95(3) $&$ 1.28(3) $&$ 1.37(3) $&$ 1.58(5) $\\
     $\Delta/(2\pi\text{GHz)} $&$ 130(10) $&$110(10) $&$120(10) $&$110(10)$&$ 90(10) $\\
     $\bar{n}_{1} $  &$0.8(1) $&$0.3(1) $&$0.6(1) $&$0.7(2) $&$0.2(1) $\\
     $\bar{n}_{2} $  &   &$1.0(2) $&$1.1(2) $&$1.0(3) $&$1.0(2) $\\
     $\bar{n}_{3} $  &   &   &$0.9(2) $&$1.0(3) $&$1.1(2) $\\
     $\bar{n}_{4} $  &   &   &   &$1.0(4) $&$0.7(2) $\\
     $\bar{n}_{5} $  &   &   &   &   &$2.3(4) $\\
    \hline\hline
  \end{tabular}
  \caption{\label{tab.params}Experimental parameters. Lowest (highest) frequency axial vibrational mode $\omega_1 (\omega_N)$, effective magnetic field $\omega_{\text z}$ varied within interval, spin coupling rate $\Omega$, detuning of the Raman beams from the $3S_{1/2}$ to $3P_{3/2}$ transition,
measured mode occupation number $\bar{n}_{j=1\ldots N}$. The detection duration is $40\;\mu$s for $N\leq2$ and $80\;\mu$s for $N\geq3$.}
\end{table}

All terms incorporating the spin in the Hamiltonian are implemented via TPSR as well. In order to achieve strong coupling, i.e., Rabi frequencies up to $\Omega/(2\pi)\approx 1.6$ MHz, 
we use intensities $I_{\text{RR}} = 2.5(5)\times 10^3I_{\text{sat}}$ and $I_{\text{BR}}\leq 0.8(1)\times 10^3I_{\text{sat}}$. We employ two acousto-optical modulators to switch on and off beams (turn coupling on/off), to tune the relative frequency difference between RR and BR, i.e., the detuning of the TPSR (varying $\omega_{\text{z}}$), and to attenuate the intensity of BR (fine tune $\Omega$). 

At intensities $I_{\text{RR}}$ and $I_{\text{BR}}$, the  contribution of  spontaneous Raman scattering~\cite{ozeri_errors_2007} of $^{25}$Mg$^+$ to residual decoherence in our experiments, is calculated to $\gamma_{\text{dec}}\leq 7(1)$ kHz, assuming a relevant beam waist ($1/e^2$ radius of intensity) $w=55(10)\;\mu$m for RR and BR. This corresponds to less than $0.07(1)$ scattering events in $t =13\tau_{\text S}$, resulting in a residual heating effect of less than $0.02$ quanta (in $13 \tau_{\text S}$) distributed among $N$ motional modes. This is in agreement with dedicated calibration measurements of the decoherence rate induced by RR and BR.
In Fig.~\ref{supp_dec_flops}, we compare typical experimental results for $\langle \sigma_{z} (t) \rangle$ to corresponding calculated unitary time evolutions with and without considering a decoherence rate $\gamma_\text{dec}$. From this we infer that $\gamma_\text{dec}\tau_{\text S} \lesssim 0.01$ such that decoherence contributes to a relative uncertainty of (in most cases much less than) $6\%$ to our derived quantities during $[\tau_\text{S},13\tau_\text{S}]$. We neglect this effect throughout our manuscript and consider our system completely isolated from external baths.


\section{Applicability of microcanonical averages}
\begin{figure}[htb]
  \centering
  \includegraphics[]{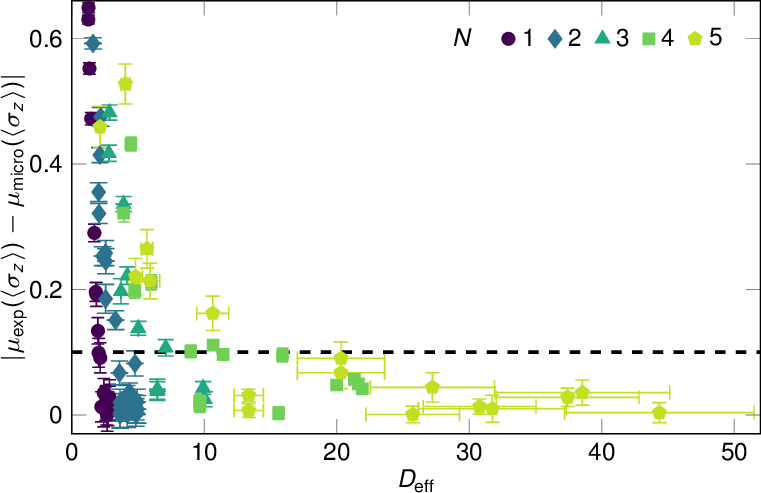}
   \caption{\label{supp_deff_deth}Comparison of experimental results to microcanonical averages. Deviation from ${\mu_{\text{micro}}}(\langle \sigma_z \rangle)$ (with statistical uncertainties) as a function of $D_{\text{eff}}$ (with systematic uncertainties) for $N=1\ldots 5$. We observe $1<D_{\text{eff}}<50$, while ranges of experimentally accessible values of $D_{\text{eff}}$ for fixed $N$ overlap, and, additionally, grow with increasing $N$. At fixed $N$, ${\mu_{\text{exp}}}(\langle \sigma_z \rangle)$ approaches ${\mu_{\text{micro}}}(\langle \sigma_z \rangle)$ with increasing $D_{\text{eff}}$ and the points with a deviation $\leq 0.1$ (dashed line) are used for evaluation in Fig.~4. 
}
 \end{figure}
In Figure 3(a), we display time averages $\mu_{\text{exp}}(\langle \sigma_z\rangle)$ as a function of effective magnetic field $\omega_{\text z}$ for $N=1\ldots5$ and compare them with the microcanonical ensemble averages $\mu_{\text{micro}}(\langle \sigma_z\rangle)$. We correlate the deviation $|\mu_{\text{exp}}(\langle \sigma_z\rangle) - \mu_{\text{micro}}(\langle \sigma_z\rangle)|$ with the corresponding effective dimension $D_{\text{eff}}$ for each data point, see Fig.~\ref{supp_deff_deth}. For this sampling of $\omega_{\text z}$, we find $D_{\text{eff}}$ lying between $1$ and $50$, e.g., for $N=2$ it can be tuned from $2$ to $5$, while for $N=5$ the spread is $2$ to $50$. This implies that we can explicitly compare experiments which have equal $D_{\text{eff}}$ despite distinct values of $N$. Further, Fig.~\ref{supp_deff_deth}, and accompanying numerical calculations, suggest that we can continuously scale $D_{\text{eff}}$ by increasing $N$ and, particularly, tuning $\omega_{\text z}$. Our definition of $D_{\text{eff}}$ measures the ability of our system to thermalize, in the sense that $\mu_{\text{exp}}(\langle \sigma_z\rangle )$ approaches $\mu_{\text{micro}}(\langle \sigma_z \rangle )$ as $D_{\text{eff}}$ increases. Based on this, we select the data points for which we study the correlations between fluctuations and effective dimension in Fig.~4: The dashed line in Fig.~\ref{supp_deff_deth} indicates a deviation of $0.1$, all measurements closer to $\mu_{\text{micro}}(\langle\sigma_z\rangle)$ are taken into account in Fig.~4.

\section{Statistical uncertainties of time averages and of mean amplitudes of time fluctuations}
In Figs.~3 and 4, we show time averages, ${\mu_{\text{exp}}}(\langle \sigma_z \rangle)$, and mean amplitudes of time fluctuations, ${\delta_{\text{exp}}}(\langle \sigma_z \rangle)$, both derived from experimentally measured time evolutions, $\langle \sigma_z(t) \rangle$. In the following, we discuss how we obtain statistical uncertainties for both quantities using the example of ${\delta_{\text{exp}}}(\langle \sigma_z \rangle)$.
The quantity ${\delta_{\text{exp}}}(\langle \sigma_z \rangle)$, as defined in Eq.~(7), represents a measure of the mean amplitude of time fluctuations. That is, ${\delta_{\text{exp}}}(\langle \sigma_z \rangle)$ itself is the standard deviation of the expectation values $\langle \sigma_z(t) \rangle$ from their time averaged mean value ${\mu_{\text{exp}}}(\langle \sigma_z \rangle)$. The statistical uncertainty of ${\delta_{\text{exp}}}(\langle \sigma_z \rangle)$, i.e., its standard deviation, can be derived from the particular form of the underlying probability distribution. To estimate this, we use a bootstrapping method: For each experimentally measured time evolution ($\sim100$ data points), we compile 100,000 data sets using the same number of data points by sampling with replacement. From these, we derive the mean value of ${\delta_{\text{exp}}}(\langle \sigma_z \rangle)$ and its standard deviation. The latter is then used as statistical uncertainty (error bar) for ${\delta_{\text{exp}}}(\langle \sigma_z \rangle)$.

\section{Eigenstate Thermalization Hypothesis and mean amplitudes of time fluctuations}
In the following, we give a heuristic derivation of the scaling of mean amplitudes of time fluctuations with the effective dimension, see Eq.~\eqref{first.final.relation},  as observed in our experiment. For this purpose, we mathematically formulate  the ETH and explain it in detail.

We consider time fluctuations of an operator $O$. We define mean amplitudes of time fluctuations ${\delta_{\infty}}(\langle O \rangle)$ (see  main text), corresponding to the fluctuations of $\langle O \rangle$ averaged over infinite time. 
If the system is initially in a pure state 
$| \phi_\alpha \rangle$ and if it has nondegenerate energy gaps,  we get~\cite{srednicki_approach_1999}
\begin{equation}
	{\delta_{\infty}}^2(\langle O \rangle) = \sum_{\genfrac{}{}{0pt}{}{\beta_1,\beta_2}{\beta_1 \neq \beta_2}} 
	|c_{\beta_1}(\alpha)|^2  |c_{\beta_2}(\alpha)|^2 |O_{\beta_1,\beta_2}|^2,
\label{time.fluctuations}
\end{equation} 
which depends on both, the structure of the many-body eigenfunctions $c_{\beta}(\alpha)$ and the matrix elements $O_{\beta_1,\beta_2}$. 

We start by discussing the properties of $c_{\beta}(\alpha)$. Motivated by our experimental conditions, we consider that initial states are  product states of the form 
$| \phi_{\alpha} \rangle = | s \rangle_z | n_1 \rangle_1 \dots |n_N \rangle_N$,
where $|s\rangle_z$ is a spin state in the $z$-basis, and $|n_j\rangle_j$ is a Fock state with $n_j$ phonons in the $m$th vibrational mode.
We can write our Hamiltonian as $H = H_0 + H_{\text I}$, where 
\begin{eqnarray}
H_0 &=& \frac{\hbar \omega_z}{2} \sigma_z + \sum_j \hbar \omega_j a^\dagger_j a_j,
\nonumber \\
H_{\text I} &=& 
\frac{\hbar \Omega}{2} \sigma_x +  
\frac{\hbar \Omega}{2} (\sigma^+ C + \sigma^- C^\dagger) .
\end{eqnarray} 
The interaction term $H_{\text I}$ (which includes both the spin coupling and the spin-phonon coupling term of the spin Hamiltonian) couples any given initial state $|\phi_\alpha \rangle$ to different states $|\phi_{\alpha'}\rangle$, within an energy shell of certain width $W_\alpha$.
Thus, we can expect that the many-body eigenstates, $| \psi_\beta \rangle$, will participate in the noninteracting initial state only within a range of energies 
$E_\beta = \bar{E}_\alpha \pm W_\alpha$. Here, $\bar{E}_\alpha$ and $W_\alpha$ are the mean energy and standard deviation of $H$ with respect to $|\psi_\alpha\rangle$, respectively. This observation motivates the ansatz
\begin{equation}
c_{\beta}(\alpha) = F(E_\beta,\alpha) C_{\beta,\alpha},
\label{ass.C}
\end{equation}
where $F(E_\beta,\alpha)$ is a smooth function of $E_\beta$ of width $W_\alpha$ defining the energy shell. The function  $F(E_\beta,\alpha)$ satisfies the normalization condition 
$\int D(E_\beta) |F(E_\beta,\alpha)|^2 d E_\beta = 1$, where we introduce the density of states, $D(E) = \sum_\beta \delta(E - E_\beta)$. Note, $D(E)$ is not representing the effective dimension, but we will relate the two in the following, see Eq.~\eqref{ipr.density}. 
The particular details of $H_{\text I}$ will induce a factor that is modeled in our ansatz by random variables, $C_{\alpha,\beta}$, with average, $\overline{|C_{\alpha,\beta}|^2} = 1$. 
This ansatz is further motivated by random matrix theory: if the interaction Hamiltonian, $H_{\text I}$, is a random matrix perturbation, it can be shown that the many-body eigenstates satisfy the energy shell condition~\cite{casati_band-random-matrix_1993}. 
We expect a  similar behavior to occur in many-body eigenstates of nonintegrable systems. Our numerical calculations qualitatively confirm this assumption. The wavefunctions appear localized in energy-space, as assumed in the energy-shell model, with a superimposed random component. We find numerically that the energy width takes the value $W_\alpha \approx \hbar \Omega$ for all initial states of the form $| \phi_\alpha \rangle = \sum_{\beta} c_\beta(\alpha)| \psi_\beta \rangle$.

The ETH conjecture explains the mechanism behind thermalization of nonintegrable closed quantum systems by assuming that thermalization is a property of single eigenstates. 
The meaning of this statement can be expressed as a condition fulfilled by matrix elements of an operator $O$ in the eigenstate basis of the nonintegrable Hamiltonian~\cite{srednicki_approach_1999}, 
\begin{eqnarray}
O_{\beta_1,\beta_2} 
&=& \langle \psi_{\beta_1} | O | \psi_{\beta_2} \rangle\\
&=&{\mathcal O}(E) \ \delta_{\beta_1,\beta_2} + \frac{1}{\sqrt{D(E)}} f(E,\omega) R_{\beta_1,\beta_2}.\nonumber
\label{ETH.explained}
\end{eqnarray} 
Here, $E = (E_{\beta_1} + E_{\beta_2})/2$ and  $\omega = (E_{\beta_1} - E_{\beta_2})$,
with the eigenenergies $E_{\beta}$ of our coupled system,
and ${\mathcal O}(E)$ is a smooth function of $E$, which corresponds to the microcanonical average at energy $E$. 
The second term on the r.h.s.~of Eq.~\eqref{ETH.explained} determines the nondiagonal matrix elements of $O$ and $f(E,\omega)$ is a smooth real function of $(E$, $\omega)$. It is centered around $\omega = 0$, and it has a typical width $W(E)$ as a function of $\omega$. The set of complex numbers $R_{\beta_1,\beta_2}$ is described by stochastic variables when averages are taken over the set of energy eigenstates
\begin{equation}
\sum_{\beta_1,\beta_2} \sum_{\beta'_1,\beta'_2} 
R^*_{\beta_1,\beta_2} R_{\beta'_1,\beta'_2} \propto \delta_{\beta_1,\beta_1'} \delta_{\beta_2,\beta_2'},
\label{R.stat}
\end{equation}
where the proportionality factor depends on the operator $O$.

The normalization of the function $f(E,\omega)$ in Eq.~\eqref{ETH.explained} requires some additional attention. 
We consider first the following summation over matrix elements:
\begin{eqnarray}
\sum_{\beta_2} |O_{\beta_1,\beta_2}|^2 &=&
\sum_{\beta_2} 
\langle \psi_{\beta_1} | O | \psi_{\beta_2} \rangle
\langle \psi_{\beta_2} | O | \psi_{\beta_1} \rangle\nonumber\\ 
&=&\langle \psi_{\beta_1} | O^2 | \psi_{\beta_1} \rangle = 1,
\label{normalization}
\end{eqnarray}
On the other hand, using Eq.~\eqref{ETH.explained}, we find
\begin{eqnarray}
\sum_{\beta_2} |O_{\beta_1,\beta_2}|^2&=& \\
O_{\beta_1,\beta_2}^2 &+& \sum_{\beta_2 (\neq \beta_1)} \frac{1}{D(E)} |f(E,\omega)|^2 
R^*_{\beta_1,\beta_2} R_{\beta_1,\beta_2}.\nonumber
\end{eqnarray}
According to the ETH, $O^2_{\beta_1,\beta_2}$ is the squared microcanonical average of operator $O$ for initial states with mean energy $E$. However, focusing on the case in which $O$ is a spin operator, $\sigma_l\;(l=x,y,z)$, we note that,
 in the range of parameters in which we observe thermalization, we detect small mean values of $\sigma_z$, and this contribution can be neglected. Otherwise, it would lead to a smooth dependence of the normalization condition on the variable $E$, which could be incorporated into the discussion below.
Using the statistical properties of $R_{\beta_1,\beta_2}$, we get:
%
\begin{eqnarray}
\sum_{\beta_2}|O_{\beta_1,\beta_2}|^2 
&\approx& \sum_{\beta_2(\neq \beta_1)} \frac{1}{D(E)} |f(E,\omega)|^2 
R^*_{\beta_1,\beta_2} R_{\beta_1,\beta_2}\nonumber\\ 
&\approx& \int dE_{\beta_2} D(E_{\beta_2}) \frac{1}{D(E)} |f(E,\omega)|^2 .
\end{eqnarray}
%
Next, we use the approximation $D(E_{\beta_2}) = D(E - \omega/2) \approx D(E)$. This is valid as long as $D(E)$ is a smooth function that satisfies 
$D''(E) W(E)^2 / D(E) \ll 1$, such that we can substitute  $D(E_{\beta_2})$ by its average value $D(E)$. Our numerical calculations show that this approximation is valid within the range of parameters considered in this work.
Finally, using this approximation together with the normalization condition given by Eq.~\eqref{normalization}, we find
\begin{equation}
1 \approx \int d E_{\beta_2} D(E_{\beta_2}) \frac{1}{D(E)} |f(E,\omega)|^2 \approx
\int d \omega |f(E,\omega)|^2 .
\end{equation}
We can also estimate the width of $f(E,\omega)$ in terms of the typical width of the energy shell, $W_\alpha$. For this, let us write explicitly the matrix elements of our reference observable, $\sigma_z$, in terms of the unperturbed basis of states, $|\phi_\alpha \rangle$,
\begin{equation}
(\sigma_z)_{\beta_1,\beta_2} = \sum_{\alpha_1} c_{\beta_1}(\alpha_1)^* 
\langle \phi_{\alpha_1} | \sigma_z | \phi_{\alpha_1} \rangle c_{\beta_2}(\alpha_1),
\label{width.f}
\end{equation}
where we have used explicitly the fact that $\sigma_z$ is diagonal in the $|\phi_\alpha\rangle$ basis. 
Since $c_{\beta_1}(\alpha_1)$, $c_{\beta_2}(\alpha_1)$ are functions of $E_{\beta_1}$, $E_{\beta_2}$, of width $W_{\alpha_1}$, the energy width of $(\sigma_z)_{\beta_1,\beta_2}$ will be determined by the typical energy width, $W_{\alpha_1}$ of eigenstates $|\phi_{\alpha_1}\rangle$ in the sum. We can even predict from Eq.~\eqref{width.f} that $W(E) \approx 2 W_{\alpha} \approx 2 \hbar \Omega$.

Let us use the estimations above in the calculation of mean amplitudes of time fluctuations. The latter can be expressed, after using the ETH Eq.~\eqref{ETH.explained} and the energy shell ansatz (Eq.~\eqref{ass.C}), like
\begin{eqnarray}
	{\delta_{\infty}}^2(\langle \sigma_z \rangle) &\approx&
	\int d E_{\beta_1} d E_{\beta_2} 
	D(E_{\beta_1}) D(E_{\beta_2}) \times\nonumber\\
	&\times&|F(E_{\beta_1},\alpha)|^2  |F(E_{\beta_2},\alpha)|^2\times\nonumber\\
	&\times&\frac{1}{D(E)} |f(E,E_{\beta_1} - E_{\beta_2})|^2 .
	\label{var.detail1}
\end{eqnarray}
The key observation to evaluate this integral is that both $F(E_\beta,\alpha)$ and $f(E,\omega)$, have a similar energy width in $E_\beta$ and $\omega$, respectively, which corresponds to the typical energy width of the energy shell, $W_\alpha$. Using the normalization condition for those functions, we can estimate,
\begin{equation}
	{\delta_{\infty}}^2(\langle \sigma_z \rangle) \propto \frac{1}{D(\bar{E}_\alpha) W_\alpha}.
	\label{var.detail2}
\end{equation}
This means that squared time fluctuations approximately decay as the inverse of the number of states within the energy shell. 
Assuming, for example, Gaussian distributions for 
$F(E_\beta,\alpha)$ and $|f(E,\omega)|^2$ of width $W_\alpha$ and $2 W_{\alpha}$, respectively, yields 
${\delta_{\infty}}(\langle \sigma_z \rangle) \approx 0.48/\sqrt{D(\bar{E}_\alpha) W_\alpha}$. However, any similar functional dependence that satisfies the normalization and energy-shell condition will yield a similar scaling with $D(\bar{E}_\alpha) W_\alpha$.

We can finally relate mean amplitudes of time fluctuations to the IPR by noting that,
\begin{eqnarray}
{\text{IPR}}(|\phi_\alpha \rangle) 
&=& \frac{1}{\sum_{\beta} |c_\beta(\alpha)|^4}\nonumber\\ &\approx& 
\frac{1}{\int_{E_\beta} D(E_\beta) |F(E_\beta,\alpha)|^4} \nonumber\\ &\propto&
D(\bar{E}_\alpha)W_\alpha . 
\label{ipr.density}
\end{eqnarray}
This relation is, again, a result of the normalization and energy shell condition of the function $F(E_\beta,\alpha)$. It has a very physical interpretation, since it relates the IPR to the number of states within the energy shell defined by $W_\alpha$. Combining the last two equations, we arrive at
\begin{equation}
{\delta_{\infty}}^2(\langle \sigma_z \rangle) \propto
\frac{1}{{\text{IPR}}(|\phi_\alpha\rangle)}.
\label{final.relation}
\end{equation}
Our numerical calculations confirm the scaling predicted by Eq.~\eqref{final.relation}. In  Fig.~\ref{fluctVsIPRsz} we calculate numerically the mean amplitudes of time fluctuations by using Eq.~\eqref{time.fluctuations} for a variety of initial conditions and system parameters. Our heuristic derivation assumes that the system is ergodic enough, so that a large number of many-body eigenstates participate in the initial state. Thus, to check Eq.~\eqref{final.relation}, we choose a wide range of parameters where the system is observed to thermalize efficiently. Figure~\ref{fluctVsIPRsz} shows that the estimate given by Eq.~\eqref{final.relation} works surprisingly well, even for small numbers of particles.

\end{document}